# Three-dimensional Reconstruction and Propagation of an Asymmetric Flux-rope Coronal Mass Ejection

**Philippe Lamy[1]●● · Yannick Boursier[2] · Jean Loirat[3] · Andrei Zhukov[4,5]**

**Abstract**
We report on the characterization of a coronal mass ejection (CME) observed on 22 October 2003 by the LASCO-C2 and C3 coronagraphs over a time interval of 6 hours. This CME clearly appears as an asymmetric flux-rope in self-similar expansion and in spite of having a single vantage point, this relatively simple morphology and the geometry of the observations allow us to reconstruct its shape and its trajectory. The analysis is based on forward modeling of an asymmetric structure where the plasma is homogeneously distributed in a thin shell and synthetic images are calculated from Thomson scattering by the electrons. They are best fitted to the images to determine the exact shape of the flux rope, and to track its evolution characterized by a radial propagation in self-similar expansion. The analysis supports the forward propagation (over the backward

✉ P. Lamy
philippe.lamy@latmos.ipsl.fr

Y. Boursier
boursier@cppm.in2p3.fr

J. Loirat
jean.loirat@gmail.com

A. Zhukov
andrei.zhukov@sidc.be

1 Laboratoire Atmosphères, Milieux et Observations Spatiales, CNRS & UVSQ, 11 Bd d'Alembert, 78280 Guyancourt, France

2 Aix-Marseille Université, CNRS/IN2P3, CPPM, Marseille, France

3 Laboratoire d'Astrophysique de Marseille, CNRS & Aix-Marseille Université, 38 rue Frédéric Joliot-Curie, 13388 Marseille, France

4 Solar-Terrestrial Centre of Excellence - SIDC, Royal Observatory of Belgium, Avenue Circulaire 3, Brussels 1180, Belgium

5 Skobeltsyn Institute of Nuclear Physics, Moscow State University, 119991 Moscow, Russia

one) characterized by constant accelerations of 76 and 46 m $s^{-2}$ for the front and rear parts of the CME, respectively. The respective velocities at 20 $R_\odot$ reach 2000 and 1100 km $s^{-1}$, and its mass unbiased by projection effects is estimated at 1.7x$10^{16}$ g. Altogether, these are quite exceptionally large values among CMEs and imply a very energetic event. No erupting event could be identified at or near the calculated initial location of the CME. It lies 25° west of the active region NOAA 10484, the site of contemporary, violent, and recurrent solar eruptions (the "Halloween event") and a putative connection would require a strong deflection over a very short path. Therefore, the origin of this CME remains unclear.

## 1. Introduction

After the discovery of coronal mass ejections (CMEs) in 1970 (Tousey et al., 1973), their three-dimensional (3D) modeling has been actively pursued to analyze and interpret the observations from white-light coronagraphs. As a first motivation, it was realized that their 3D morphology would give clues to the mechanism(s) of their formation. The bubble or shell-like geometry progressively prevailed leading to the flux rope interpretation, at least for a large fraction of them, notably those associated with filament/prominence eruptions. A second motivation is their 3D propagation, their spatial development and their interaction with the corona and the solar wind. Of particular relevance is the determination of their possible impact with the Earth, other planets, and interplanetary space probes to compare with *in situ* measurements thus contributing to space weather predictions.

A major impetus to this modeling effort was given by the Large-Angle Spectrometric COronagraph (LASCO: Brueckner et al. (1995)) aboard the *Solar and Heliospheric Observatory* (SOHO: Domingo, Fleck, and Poland (1995)) which has produced several ten thousand temporal sequences of CME ejections beginning in early 1996 and still on-going. A second one came from the SECCHI-COR2 coronagraphs (Howard et al., 2008) of the *Solar-Terrestrial Relations Observatory* (STEREO) mission launched in 2006 with one of the originally two spacecraft still active (STEREO-A).

Mierla et al. (2010) published the first comprehensive review of the techniques of 3D reconstruction of CMEs using coronagraph data, later followed by Thernisien, Vourlidas, and Howard (2011) who focused on the results of the STEREO mission. This latter paper is part of a special issue of the JASTP journal (Vol, 73, 2011) entitled "On three-dimensional aspects of CMEs, their source regions and interplanetary manifestations" showing the early interest in these questions. Let us briefly summarize the main methods of the 3D reconstruction of CMEs after noting that they depend upon the number of vantage points of observations and that they usually require several strong assumptions such as self-similar expansion as CMEs propagate outwards.

The most widely used method is the forward-modeling technique for flux-rope CMEs known as the graduated cylindrical shell (GCS) developed by Thernisien,

Howard, and Vourlidas (2006). The outer shape of such CMEs is represented by half of a torus front with two conical legs connected back to the Sun (see their Figure 1). An empirically-defined electron density function is specified in a thin shell centered on the so-defined surface thus resulting in a "hollow croissant". Synthetic images for different projections are calculated based on Thomson scattering and are compared with white-light images of CMEs. However, this model has been prominently implemented in a pure geometric approach where a wire-frame rendering is fitted to an observed CME to reconstruct its shape (e.g. Liu et al. (2010b); Feng, Inhester, and Mierla (2013); Braga et al. (2022); Patel et al. (2023); Palmerio et al. (2024); Carcaboso et al. (2024), Al-Haddad and Lugaz (2025); Singh et al. (2025)).

Few studies modeled successive observations of CMEs to determine their kinematics (speed and acceleration), notably Thernisien, Vourlidas, and Howard (2009) for 26 CMEs, Di Lorenzo et al. (2024) for 2 CMEs and Gandhi et al. (2024) for 44 CMEs. We are aware of only one case where the modeling was pursued to its ultimate phase of photometric simulation, namely the CME studied by Thernisien, Howard, and Vourlidas (2006); however, the electron density was determined only in the CME leading front.

A variant of this modeling relevant to the present investigation is the plasma cloud model proposed by Boursier, Lamy, and Llebaria (2009), basically a "hollow croissant" with no legs (i.e. unconnected to the Sun). It is defined by two concentric hemispherical surfaces (see their Figure 2) and the electron density is specified in a thin shell likewise the GCS model.

In the forward modeling technique, only the outer surface of the CME is obtained as its internal structure is not considered. It may be applied to observations from a single and multiple vantage points. Based on extensive simulations with the GCS model, Verbeke et al. (2023) showed that two vantage points significantly reduce the uncertainty in deriving CME parameters, but that the gain is marginal when further increasing the number of vantage points.

Tie-pointing and triangulation methods obviously require at least two vantage points and attempt to find a correspondence between structures in the images taken from these vantage points. This has been the method of choice for the STEREO observations of CMEs, but in practice has been limited to tracking the leading edge of CMEs (e.g. Liewer et al. (2009); Liu et al. (2010a); Liewer et al. (2011); Braga et al. (2017)).

The polarimetric reconstruction technique exploits the dependency of the degree of polarization of the Thomson scattering by electrons on the scattering angle, thus allowing the spatial localization of a small volume of coronal plasma along a given line-of-sight. This technique basically requires a single vantage point and was introduced by Poland and Munro (1976), although in a very primitive form, and later applied to LASCO images by Moran and Davila (2004), Dere, Wang, and Howard (2005), and prominently by Floyd and Lamy (2019) who reconstructed 15 CMEs. As extensively shown by these latter authors, this technique remains the only one capable of reconstructing in detail the interior of CMEs thus revealing their complex structure as recently confirmed by *Parker Solar Probe* (PSP) observations (Howard et al., 2022).

This field of research remains very active as the imaging instruments of *Solar Orbiter* (the METIS coronagraph and the SOLO heliospheric imager) and of *Parker Solar Probe* (the WISPR heliospheric imager) joined the veteran workhorses SOHO-LASCO and STEREO-A-COR2 allowing reconstructions from either multi-spacecraft observations and/or new perspectives. Recent examples in addition to the references already quoted above include: Bemporad et al. (2022), Mierla et al. (2023), Romeo et al. (2023), Yang et al. (2023), Zhuang et al. (2023), Zimbardo et al. (2023), and Stepanyuk and Kozarev (2024)).

The present investigation exploits remarkably favorable observations of a CME by the LASCO-C2 and C3 coronagraphs on 22 October 2003. The relatively simple morphology of this CME (hereafter $CME_{22}$) and the geometry of the observations allowed us to precisely define its shape and to follow its expansion and propagation over several hours. However, the modeling required the introduction of an asymmetric shape, thus generalizing past models. Another appealing aspect of $CME_{22}$ is its a priori connection to the episode of extreme solar eruptions which occurred from 18 October to 8 November 2003 (the "Halloween event"), connection which however remains unclear (Gopalswamy et al., 2005).

The article is organized as follows: In Section 2, we describe the observations and their processing. Section 3 describes the implemented 3D model of $CME_{22}$ and Section 4 how it was fitted to the observations. Section 5 presents the results for the physical properties of $CME_{22}$, the case of its mass being the topics of Section 6. The question of the source of $CME_{22}$ is considered in Section 7 in the broader context of solar activity in October 2003. In Section 8, we discuss our results and we conclude in Section 9.

## 2. LASCO Observations of the CME of 22 October 2003 and their Processing

The observations of $CME_{22}$ were acquired by the LASCO-C2 and C3 externally occulted coronagraphs and we made use of the full frame images of 1024×1024 pixels taken with the orange filter in the case of C2 and the clear filter in the case of C3 for our analysis. The orange filter has a bandpass of 540–640 nm tailored to record the continuum white-light corona and to exclude the most prominent emission lines of Fe XIV (530.3 nm) and H$\alpha$ (656.3 nm). However, it does include the He I $D_3$ line (587.7 nm, right at the center of the bandpass) emitted by cool prominence material and this is particularly relevant for the study of CMEs, notably their polarization (Floyd and Lamy, 2019). The clear filter has a very broad bandpass of 400–850 nm with negligible contributions from the above lines in the field of view of C3. All LASCO images undergo a standard processing as described in our past articles (e.g. Lamy et al. (2014); Barlyaeva, Lamy, and Llebaria (2015)) producing images fully corrected for instrumental effects and calibrated in units of $\overline{B_\odot}$. The selected images for the analysis of $CME_{22}$ were subjected to two additional operations. First, we applied the instrument vignetting functions as a radial filter to compensate for the strong gradient of the radiance of the corona and reveal its structure. Second and as standard practice in the analysis of coronagraphic images of CMEs, we subtracted a pre-event image in order to remove the background K+F corona. This assumes that the K-corona and in particular the streamers did not evolve during the observational sequence which is rarely the case, but there is unfortunately no alternative to this method. This latter operation, illustrated in Figure 1 on a C2 and a C3 images, was only partly successful as remnants of streamers spoil the CME images.

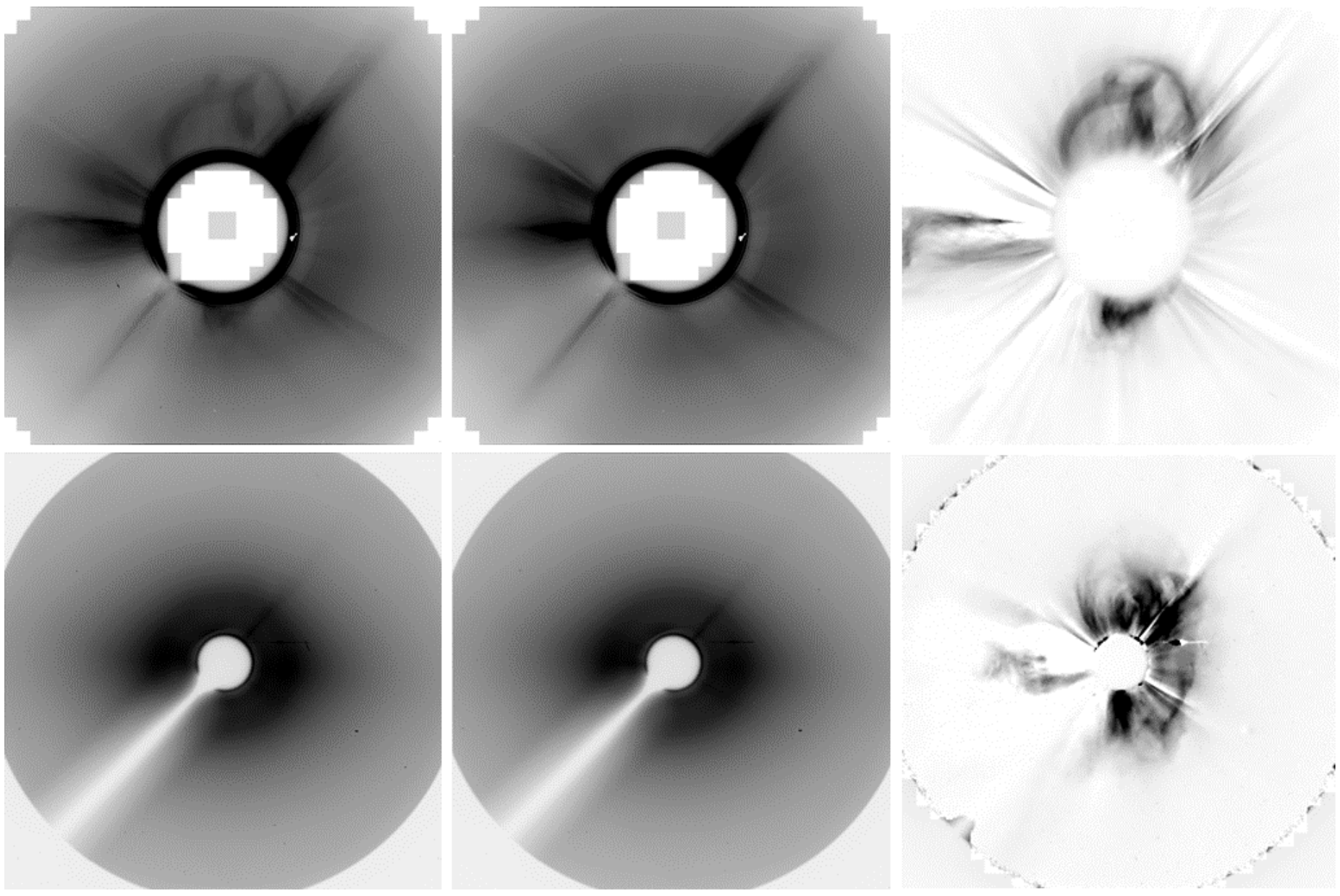

**Figure 1.** Illustration of the procedure of subtracting a pre-event image in the case of a C2 image (upper row) and a C3 image (lower row). The left column displays the original images, the central column the pre-event images, and the right column the differences. Note that in the case of C3, the original and pre-event images are nearly indistinguishable as the F-corona far exceeds the K-corona and the CME. All images are vignetted as explained in the text. In these images as well as subsequent images in the article, the projected direction of solar north is up.

$\mathrm{CME}_{22}$ became clearly visible in the C2 field of view on 22 October 2003 at 8:30 UT in the northern hemisphere at an apparent latitude of $\approx 85°$. It filled the C2 fov at 10:30 and became hardly perceptible at 11:30. The CME could be clearly observed in the C3 field of view at 9:56 and until 14:42 beyond which it vanished. Overall, $\mathrm{CME}_{22}$ could be tracked during 6 hours and 12 minutes. We selected for our analysis a set of nine C2 and nine C3 images (a total of 18 images) that sample this time interval as listed in Table 1. Figure 2 displays two views of $\mathrm{CME}_{22}$, one in the C2 field of view at 10:30 the other in the C3 field of view at 11:06. They reveal several striking features. First, the asymmetric "dry bean" shape of the CEM which suggests an asymmetric magnetic flux-rope (MFR) although there is no hint of legs anchored on the Sun on either image. A probable explanation comes from the quasi-halo morphology of $\mathrm{CME}_{22}$ which implies that it travels either toward or away from the observer. Only limb CMEs with proper MFR shape and orientation do display legs as illustrated by those of 13 April 1997 (Chen et al., 1997) and 9 September 1997 (Chen et al., 2000). Second, the resemblance of the C2 and C3 images clearly suggests a self-similar expansion of the asymmetric CME. This will be further confirmed in 4 where additional images will be displayed together with our asymmetric model. Third, the complex internal structure of $CME_{22}$ is best revealed by the associated wavelet filtered images: most prominent is a system of concentric loops forming the flux rope.

We note that a nearly contemporary CME appeared in the C2 field of view some 20 minutes before $CME_{22}$ with bright ejecta propagating radially eastward along the solar equator which are visible on the early images of the C2 sequence. In contrast, this CME does not have a well defined shape and may be considered as "Other" or "Unknown" according to the morphological classification proposed by Vourlidas et al. (2017).

**Table 1.** Journal of the selected 18 observations of the CME of 22 October 2003.

| File name | Time | Instrument | $R_f$ | $R_b$ | $\delta R$ |
|---|---|---|---|---|---|
| 22157247 | 08:30 | C2 | 2.0 | 2.1 | 0.1 |
| 22157248 | 08:54 | C2 | 2.3 | 2.4 | 0.1 |
| 22157249 | 09:06 | C2 | 2.45 | 2.5 | 0.05 |
| 22157250 | 09:30 | C2 | 2.8 | 2.9 | 0.1 |
| 22157251 | 09:54 | C2 | 3.45 | 3.6 | 0.05 |
| 22157252 | 10:06 | C2 | 3.8 | 3.9 | 0.05 |
| 22157253 | 10:30 | C2 | 4.7 | 4.9 | 0.1 |
| 22157254 | 10:54 | C2 | 5.5 | 5.9 | 0.1 |
| 22157255 | 11:06 | C2 | 6.1 | 6.6 | 0.1 |
| 32113659 | 10:20 | C3 | 4.6 | 4.8 | 0.2 |
| 32113660 | 10:42 | C3 | 5.5 | 5.9 | 0.2 |
| 32113661 | 11:18 | C3 | 7.6 | 8.2 | 0.2 |
| 32113662 | 11:42 | C3 | 8.9 | 9.5 | 0.2 |
| 32113663 | 12:18 | C3 | 10.7 | 12.0 | 0.2 |
| 32113664 | 12:42 | C3 | 12.2 | 14.2 | 0.2 |
| 32113665 | 13:42 | C3 | 15.3 | 18.8 | 0.2 |
| 32113666 | 14:18 | C3 | 17.6 | - | 0.2 |
| 32113667 | 14:42 | C3 | 19.4 | - | 0.3 |

File name: root number of the LASCO images.
Time: time of image acquisition on 22 October 2003 (UT).
Instrument: C2 or C3 coronagraphs.
$R_f$ and $R_b$: heliocentric distances of the center $C$ of the CME for the forward and backward propagation, respectively ($R_\odot$).
$\sigma$: Uncertainty on $R_f$ and $R_b$.

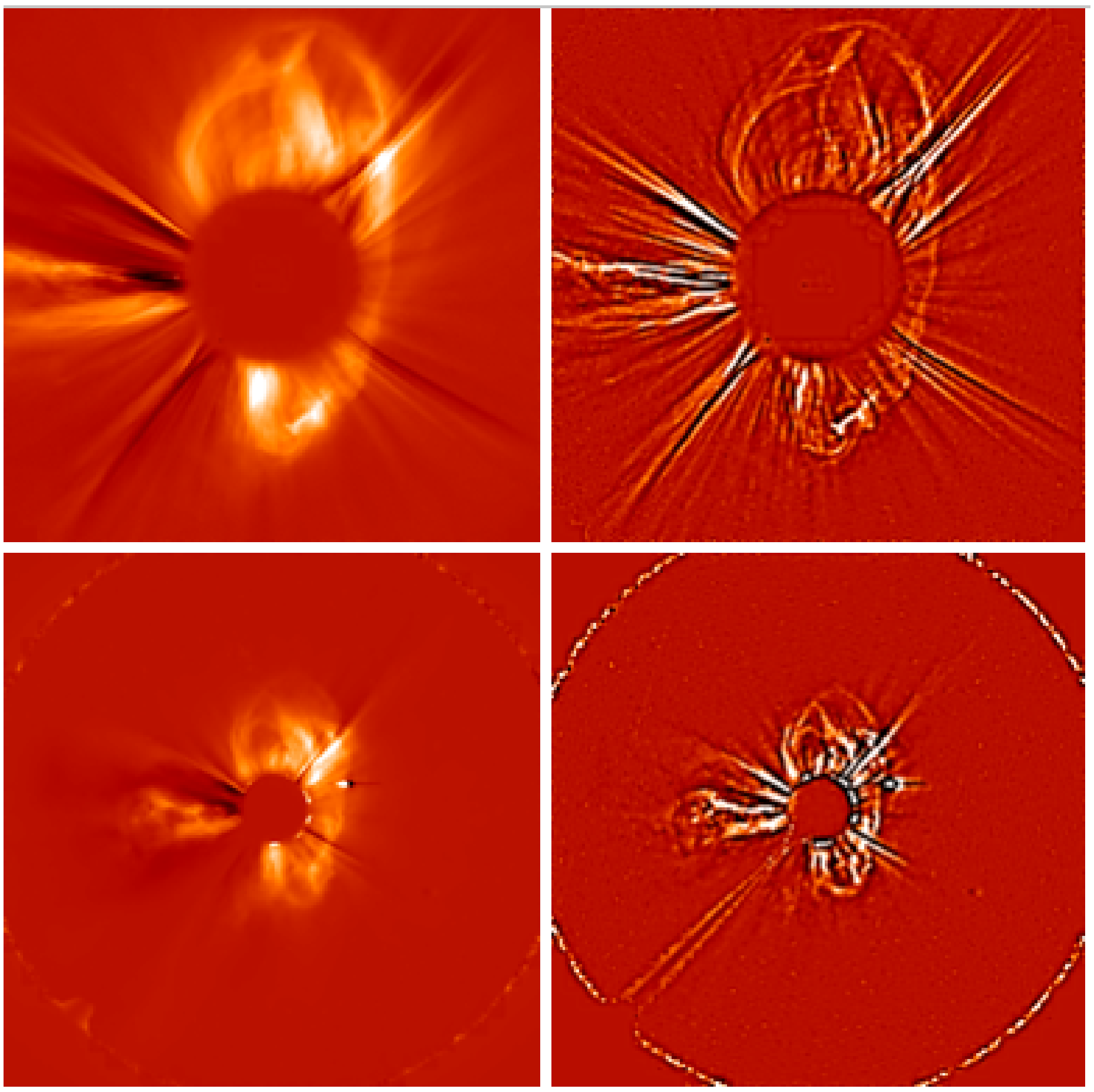

**Figure 2.** Two views of the CME of 22 October 2003 from observations taken by C2 at 10:30 (upper row) and C3 at 11:06 (lower row). The left column displays the images after application of the standard process (background subtracted and vignetting applied). In the right column, the images have been further wavelet filtered to reveal the internal structure of the CME.

## 3. The Asymmetric Plasma Cloud Model

Granting that we have a single vantage point and based on the appearance of $CME_{22}$ (Figure 2), only two models are suitable for its reconstruction, the graduated cylindrical shell and the plasma cloud, notwithstanding a substantial evolution to render them asymmetric. As explained in the introduction, the main difference between these two models lies in the presence or not of legs. There are several compelling arguments to select the latter model in the present case. First and as pointed out in the above section, there is no hint of legs on the images of $CME_{22}$ so that they cannot be constrained. Anyway, their contribution to the characterization of $CME_{22}$ (e.g. propagation, mass) remains quite marginal. Second and from the point of view of reconstruction, it is very easy to make the plasma cloud asymmetric with a limited number of parameters. Indeed, starting from the basic cloud-like model of Boursier, Lamy, and Llebaria (2009) as displayed in their Figure 2, an asymmetric plasma cloud (APC) model can be easily produced by changing the angular position of the maximal radius of the main tubular section as illustrated in Figure 3. A slight disadvantage of the APC model comes from the singularities created by the connection between the two surfaces that result in artifacts in the synthetic images. We will see later that they do not hamper the comparison with the observations.

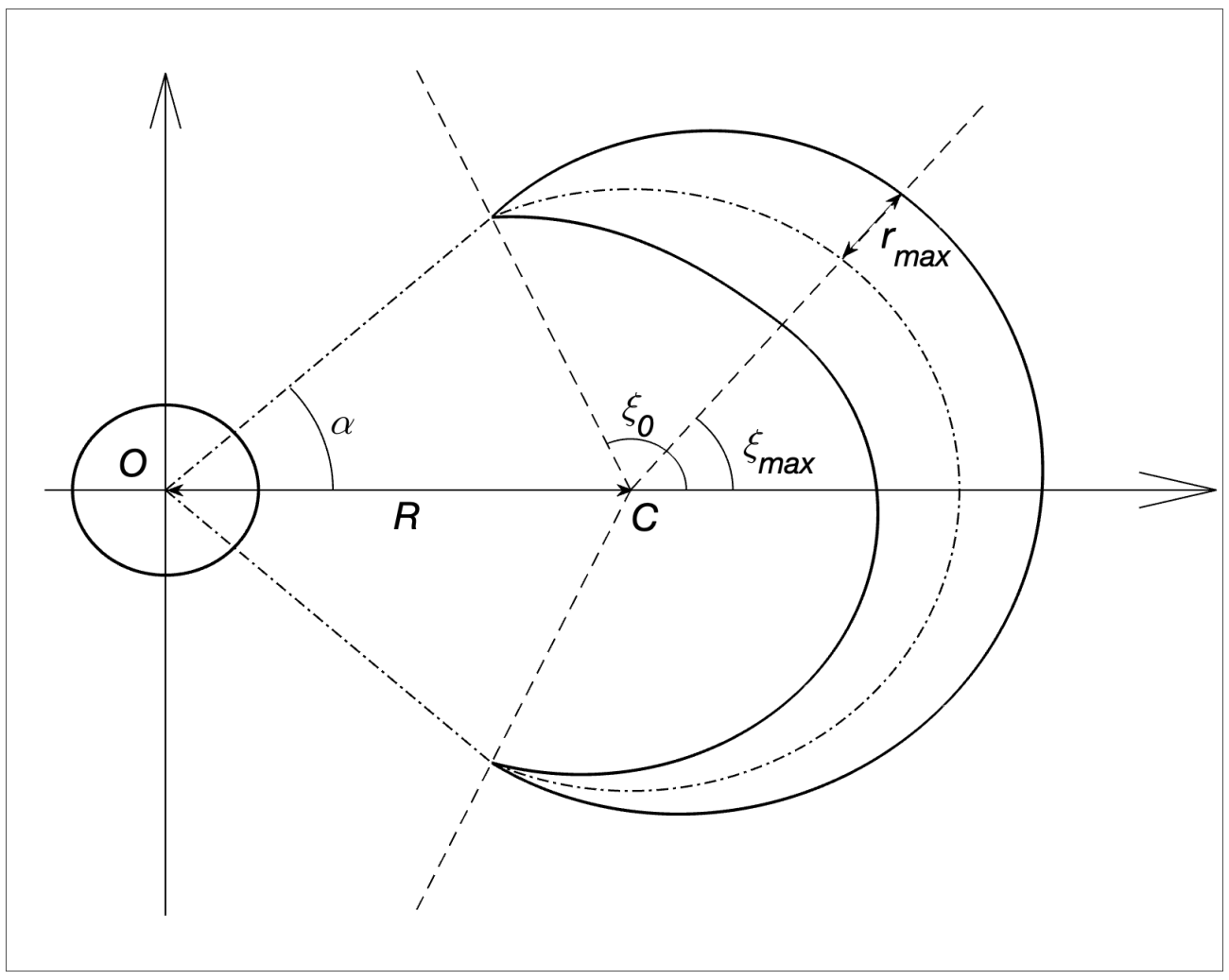

**Figure 3.** Face-on view of the asymmetric plasma cloud model. The solar disk with it center at $O$ is represented by the small circle. The center of CME$_{22}$ is defined by the points $C$ from which the tubular shell is constructed. The solid lines outline the models and they correspond to the lengthwise planar cuts through the tubular shell and point $O$. The dashed-dotted line shows the axis of the tubular shell. The end-points of the "crescent" are set by their position angle $\xi_0$ and its maximum cross-section is located by its position angle $\xi_{\max}$.

The APC model is built by first considering its cross-section in its plane of symmetry (i.e. its face-on view) as shown in Figure 3 and defining a set of parameters. The first one is the heliocentric distance $R$ of the geometric center $C$ of CME$_{22}$, that is the distance $OC$ where $O$ is the center of the Sun. The symmetric (with respect to the $OC$ axis) end-points of the "crescent" are defined by two angles $\alpha$ and $\xi_0$ with summits at $O$ and at $C$, respectively. A circle centered at $C$ and connecting the two end-points then defines the axis of the "crescent" from which its variable circular cross-section is elaborated. For any radial direction with origin at $C$ and position angle $\xi$, the circular cross-section of the "crescent" has a radius $r(\xi)$ varying between 0 at the end-points and a maximum value $r_{\max}$ reached at a position angle $\xi_{\max}$ called the asymmetry angle. The APC model is then defined by the following equations.

$$r(\xi) = \begin{cases} r_{\max}\cos(\frac{\pi}{2}\frac{\xi-\xi_{\max}}{\xi_0-\xi_{\max}}) & if \quad \xi_{\max} \le \xi \le \xi_0 \\ r_{\max}\cos(\frac{\pi}{2}\frac{\xi-\xi_{\max}}{\xi_0+\xi_{\max}}) & if \quad -\xi_0 \le \xi \le \xi_{\max} \end{cases}$$

To ensure the assumed self-similar expansion of $\mathrm{CME}_{22}$, $r_{\max}$ is proportional to the distance $R$ via a constant $\kappa$ so that:

$$r_{\max} = \kappa R \tag{1}$$

The spatial location of the model is set by the Carrington longitude and heliographic latitude of its center and by the orientation of its symmetry plane with respect to the solar equatorial plane given by the angle $\Psi$ .

Finally, we construct the shell centered on the surface of the hollow "crescent" by specifying the electron density $N_e$ within the shell. Following Thernisien, Howard, and Vourlidas (2006) and as illustrated by their Figure 1, we implement a Gaussian profile such that for any circular cross-section of the "crescent" defined by $r(\xi)$, the density profile centered on this circle falls off with Gaussian width $\sigma$ on each side.

$$N_e(d) = N_{e,0}\, exp\left(-\left(\frac{d - r_{\max}}{\sigma}\right)^2\right) \tag{2}$$

where $d$ is the radial distance in the cross-section (same notation as used by Thernisien, Howard, and Vourlidas (2006)). In our case, the asymmetric Gaussian profile used by Thernisien, Howard, and Vourlidas (2006) is not warranted as $\mathrm{CME}_{22}$ does any exhibit a well defined front. Therefore, we impose a constant width $\sigma = 0.1\ \mathrm{R}_{\odot}$. At this stage, the electron density $N_{e,0}$ is arbitrarily set.

The 3D construction of the numerical model and of the 18 synthetic images corresponding to the 18 observations listed in Table 1 relies on a method first introduced by Saez, Lamy, and Llebaria (2006) for the simulation of the streamer belt. A follow-on article by Saez et al. (2007) offers a detailed description of this method as well as applications to other large scale structures of the solar corona one of them being precisely a flux-rope CME, see their Figure 22. Briefly, the procedure involve three main steps.

- Computation of the 3D spatial distribution of electrons in $\mathrm{CME}_{22}$ thanks to the analytical expression of the model.
- Octree compression in order to efficiently reduce the computation time and the large size of data. It implements an adaptive spatial resolution allowing subdivisions proportional to the gradient of the electron density.
- Generation of synthetic images by a ray-tracing through the octree, implementing the Thomson scattering and introducing the instrumental parameters.

## 4. Comparison of the APC model with the Observations

The morphological parameters of $\mathrm{CME}_{22}$ and its trajectory are determined by a trial and error process of experimenting with different parameters to see which parameters lead to synthetic images that visually best match the real data. The fit of the model for each of the 18 observations and thus the determination of the parameters proceed as follows. In a first step, a symmetric configuration of the model is built by imposing $\xi_{\max} = 0$ and is approximately fitted to the selected image of $\mathrm{CME}_{22}$. This yields the longitude, latitude, orientation, and the angles $\alpha$ and $\xi_0$. In a second step, an asymmetric configuration of the model is built and fitted to $\mathrm{CME}_{22}$ allowing the determination of the angle $\xi_{\max}$ and the constant $\kappa$ together with $r_{\max}$. In a third step, the whole set of parameters is finely tuned for the 18 observations to produce the best match between the real and the synthetic images of $\mathrm{CME}_{22}$. All parameters, except $r_{\max}$ and $R$ which describe the dynamics of $\mathrm{CME}_{22}$ over time, turn out to be constant throughout the observations.

The procedure leads to the following determinations: a latitude of $4^\circ \pm 1^\circ$, an orientation of $90^\circ$ (i.e. along the direction of the poles), $\alpha$=45°, $\xi_0$=30°, and $\kappa$=0.5. Figure 4 shows two examples where the model is superimposed on the observations in two different displays. First, only the contours of the model are superimposed on the images demonstrating its ability to closely match the outer shape (the envelope) of $\mathrm{CME}_{22}$. Second, the synthetic images created from the model (photometrically unscaled) are superimposed on the images. The two opposite V-shaped artifacts result from the sharp extremities of the crescent as expected.

At this stage, the direction of propagation forward (that is toward Earth) or backward (that is away from Earth) remains open and the two cases are considered leading to two different solutions having longitudes of 5° and 175°, respectively. At a given heliocentric distance, the two synthetic images corresponding to the two solutions are quasi similar, differing only by their slightly different sizes as illustrated in Figure 5. It will turn out in the course of this study that the forward solution is the most likely based on several evidences. Consequently, we limit below the illustrative figures to this solution however presenting the results for the two solutions.

## 5. Properties and Propagation of the CME of 22 October 2003

Figures 6 and 7 display time sequences where four successive images of C2 and C3, respectively are directly compared with their corresponding synthetic images constructed from the APC model. These sequences clearly confirm the two assumptions of the development of $\mathrm{CME}_{22}$: radial propagation, that is along the fixed *OC* line, and self-similar expansion. No detectable deviation is observed, neither deflection nor rotation.

Having ascertained the validity of the APC model and characterized, we constructed height–time plots of different points of $\mathrm{CME}_{22}$ along the *OC* direction of propagation and the case of its center $C$ is presented in Figure 8. A first feature is the marked difference in the early phases of the propagation between the two solutions as the profiles level off at $1\,\mathrm{R}_\odot$for the forward case, but at $2\,\mathrm{R}_\odot$ for the backward one. This offers a first argument favoring the former solution. A second feature is the fact that all data sets can be very well fitted by quadratic functions, thus implying constant accelerations. The corresponding velocity profiles are displayed in Figure 9 and they confirm that the different parts of $CME_{22}$ travel as expected from the geometric model in self-similar expansion with multiplicative factors proportional to the heliocentric distance, reflecting the affine transform between successive images. A striking feature is the exceptionally large values, especially in the case of backward propagation, where the CME front reaches speed in excess of 2000 km $s^{-1}$ beyond 13 $R_\odot$ well above values usually observed Lamy et al. (2019). In our opinion, this offers a second argument favoring the forward solution for which the speed of the front reaches only 2000 km $s^{-1}$ at 20 $R_\odot$. The corresponding accelerations are given in Table 2 and here gain, the values for the backward solution are excessively large offering a third argument favoring the forward solution.

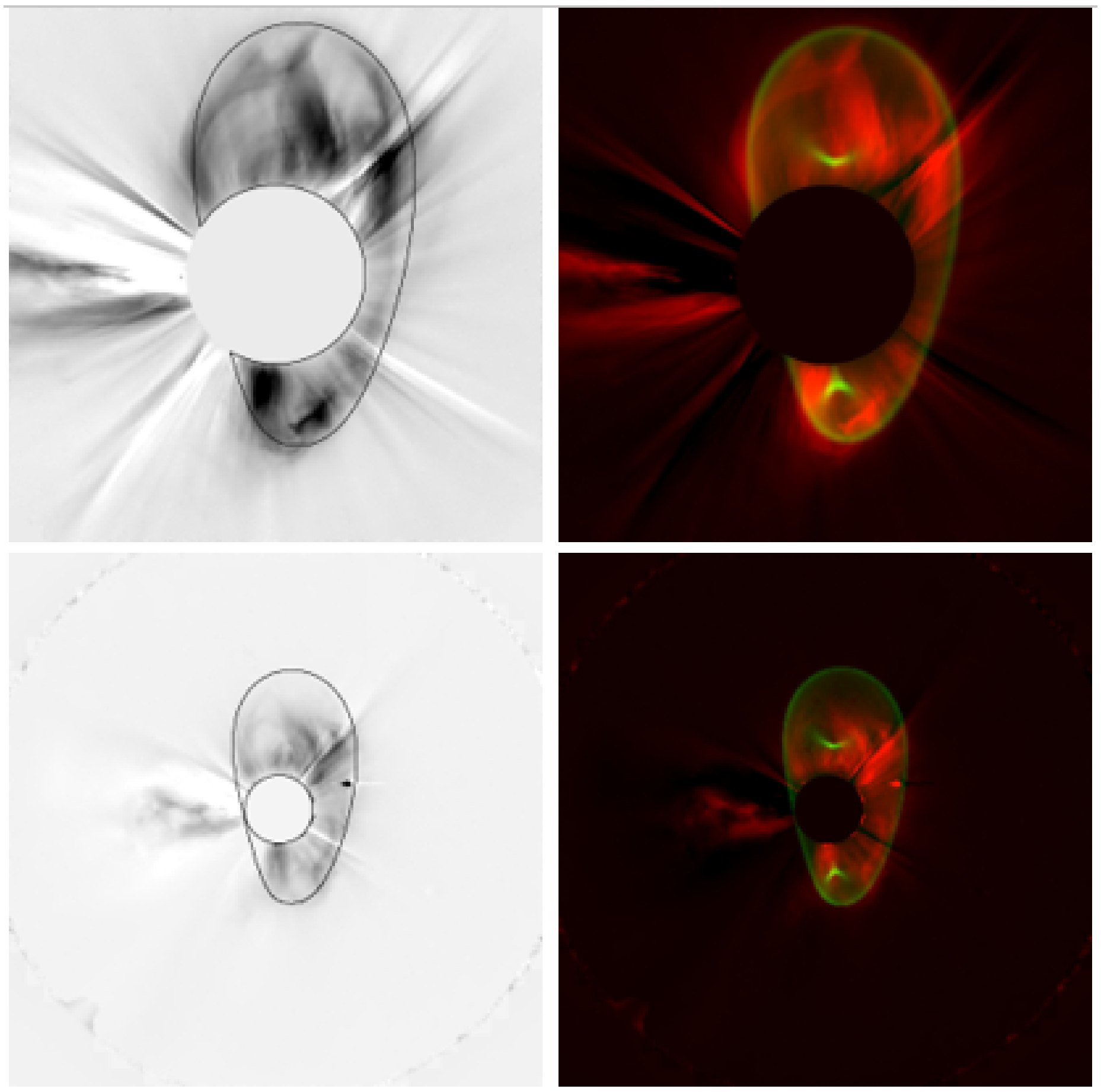

**Figure 4.** Examples of superposition of the model on $CME_{22}$ images in the case of C2 (upper row) and C3 (lower row). The left column displays only the contour of the model whereas the right column displays the synthetic images created by the model (in green). The C2 and C3 images are displayed with the same photometric scale.

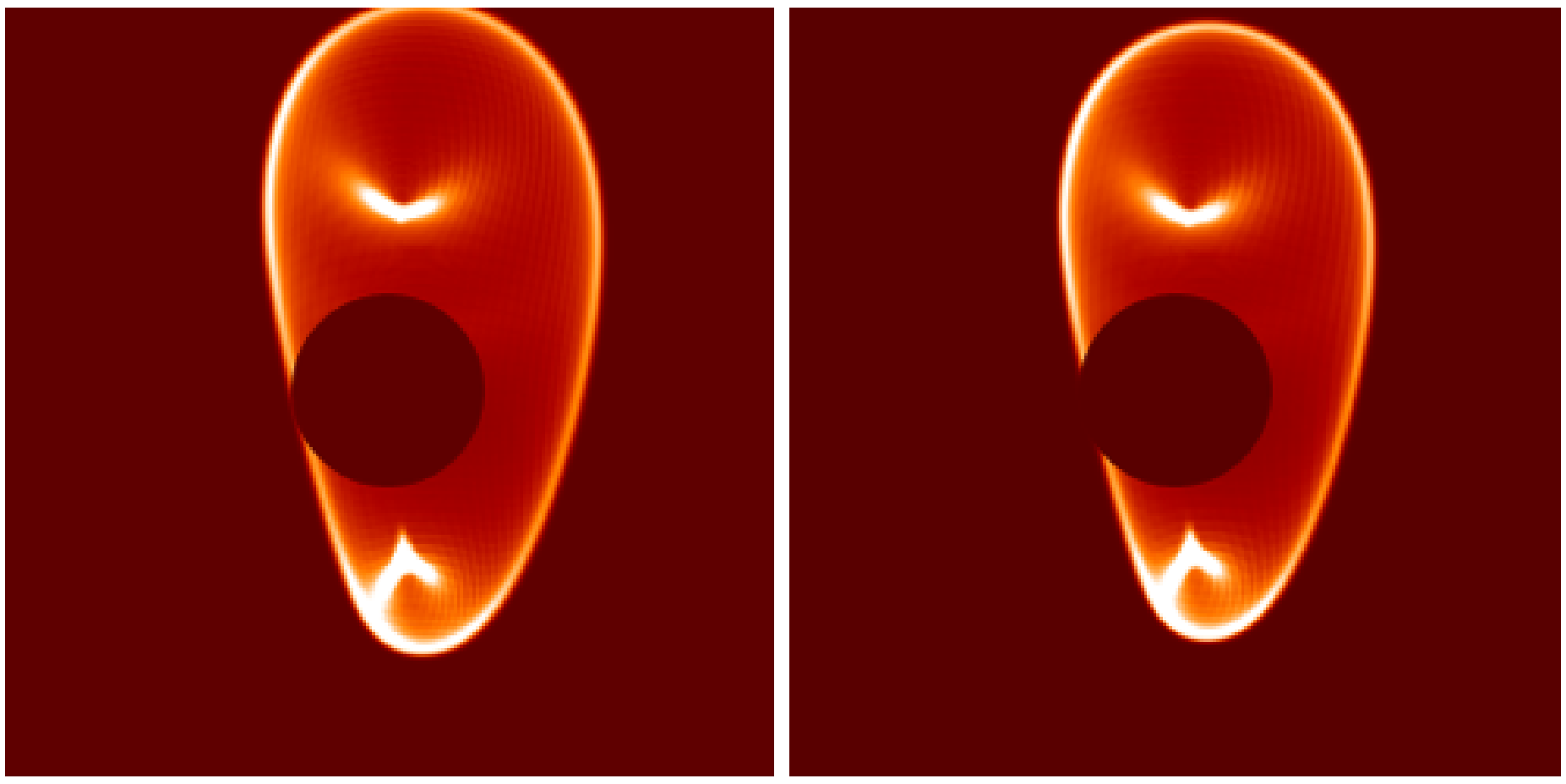

**Figure 5.** Synthetic images created by the model at an heliocentric distance of $5\,R_\odot$ corresponding to forward (left panel) and backward (right panel) propagations.

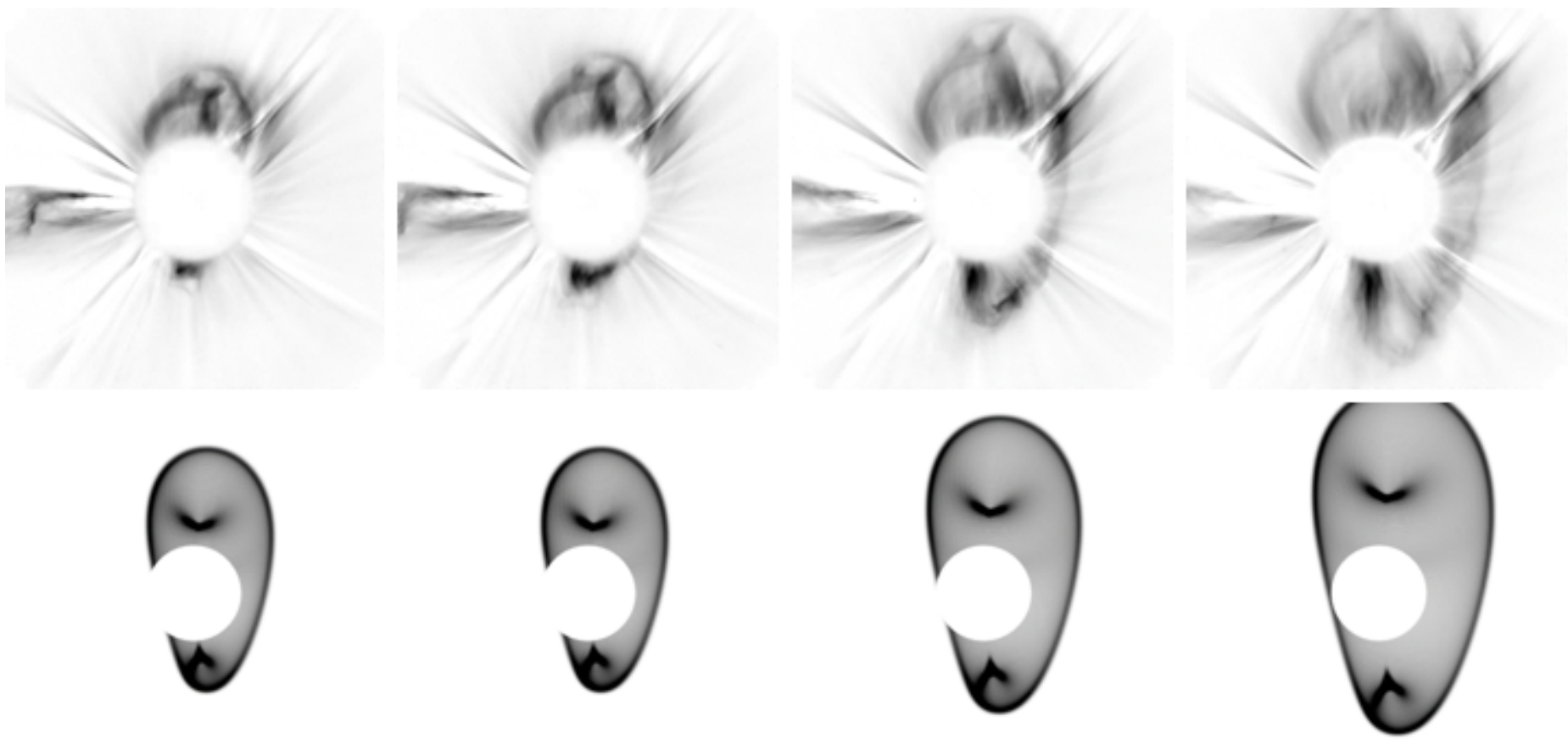

**Figure 6.** Time sequence of the propagation of $CME_{22}$ in the C2 field of view. The C2 observations on 22 October 2003 at 9:54, 10:06, 10:30, and 10:54 UT (upper row) are compared with the synthetic images from the APC model (lower row).

We now compare our results with those listed in four catalogs of CME, usually a single value relevant to the fastest or the brightest parts of the CME, implementing different methods of determination. In the present case, this is further complicated by the unfavorable geometry, $CME_{22}$ propagating at a large angle from the plane of the sky so that the projected velocities are severely biased. As a last preliminary remark, these catalogs rely on either C2 images only or on C2 + C3 images. We briefly comment their results presented in Table 3.

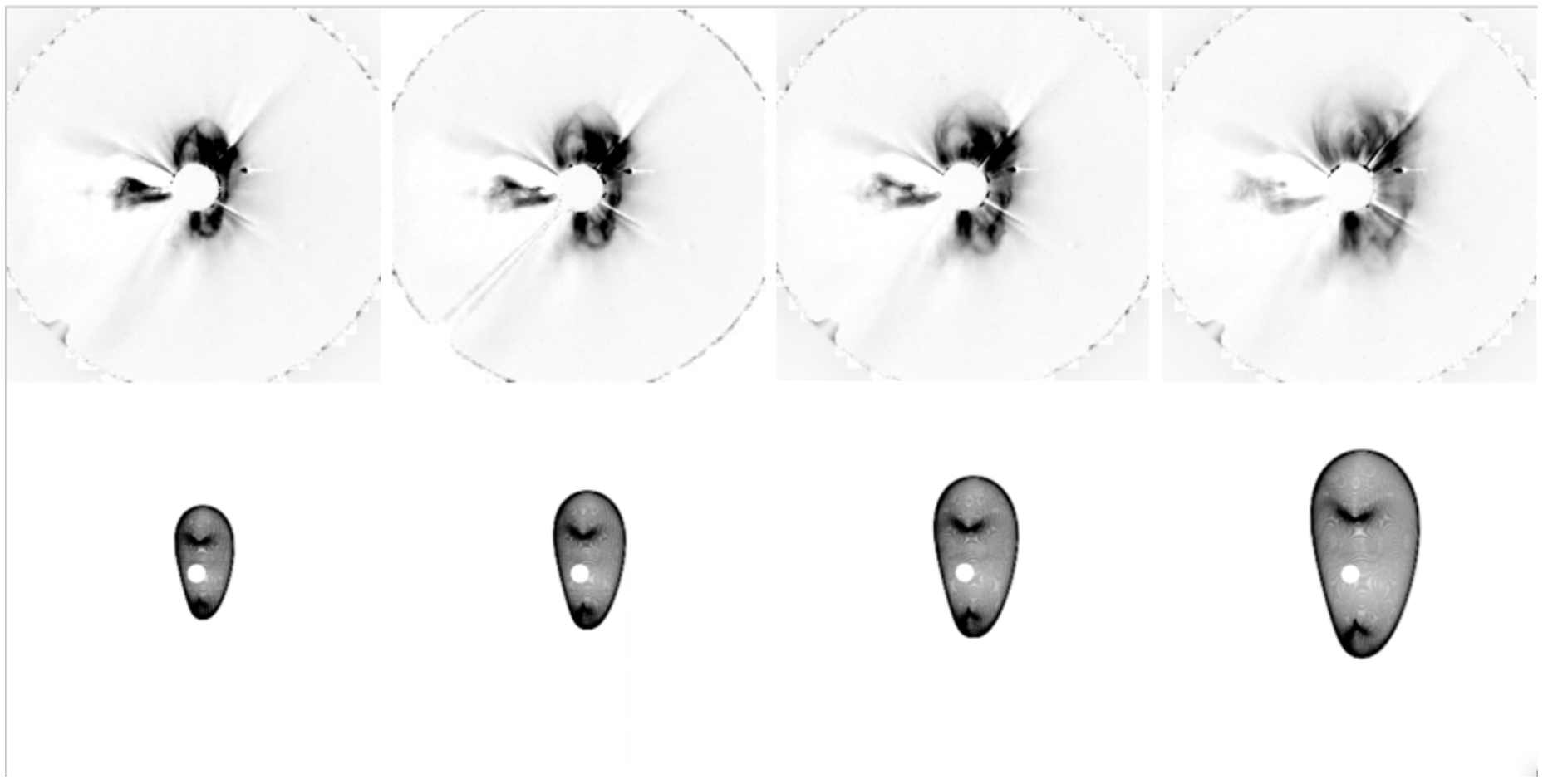

**Figure 7.** Time sequence of the propagation of $CME_{22}$ in the C3 field of view. The C3 observations on 22 October 2003 at 11:42, 12:18, 12:42, and 13:42 UT (upper row) are compared with the synthetic images from the APC model (lower row).

Considering the ARTEMIS catalog[1], we retain the "global" velocity averaged between 3 and $5.5\,R_\odot$ which gives a higher weight to the brightest parts, that is the front and central parts (which are the fastest). CACTus[2] measures a linear speed profile as a function of the position angle over the CME angular width and lists the median value. The CDAW[3] linear speed is obtained by fitting a straight line to the height–time measurements made at the fastest section of CMEs. Its 2nd order speed at $20\,R_\odot$is obtained by fitting a quadratic function. SEEDS[4] detects the northern and southern parts of $CME_{22}$ as two distinct CMEs. Their speed is taken from the highest peak using the leading-edge segmentation. We note the very large dispersion in the reported speeds and that only ARTEMIS and CDAW give large values consistent with our present results. CDAW further reports an acceleration from their quadratic fit which is roughly consistent with our determination.

We complete the picture of the forward propagation of $CME_{22}$ by estimating its launch time and initial velocity. As mentioned above, the quadratic fit to the height–time plot of the forward solution levels off at $1\,R_\odot$ making the determination of the launch time of approximately 7:00 UT rather imprecise. However, the first four data point of the height–time plot are well aligned indicating a constant velocity in the early phase of propagation. Extrapolating to $1\,R_\odot$ yields a launch time of 7:15 $\pm$ 13 minutes and an initial velocity of 154 $\pm$ 26 km $s^{-1}$.

[1] http://idoc-lasco.ias.u-psud.fr

[2] http://sidc.oma.be/cactus/

[3] http://cdaw.gsfc.nasa.gov/CME-list/index.html

[4] http://spaceweather.gmu.edu/seeds/

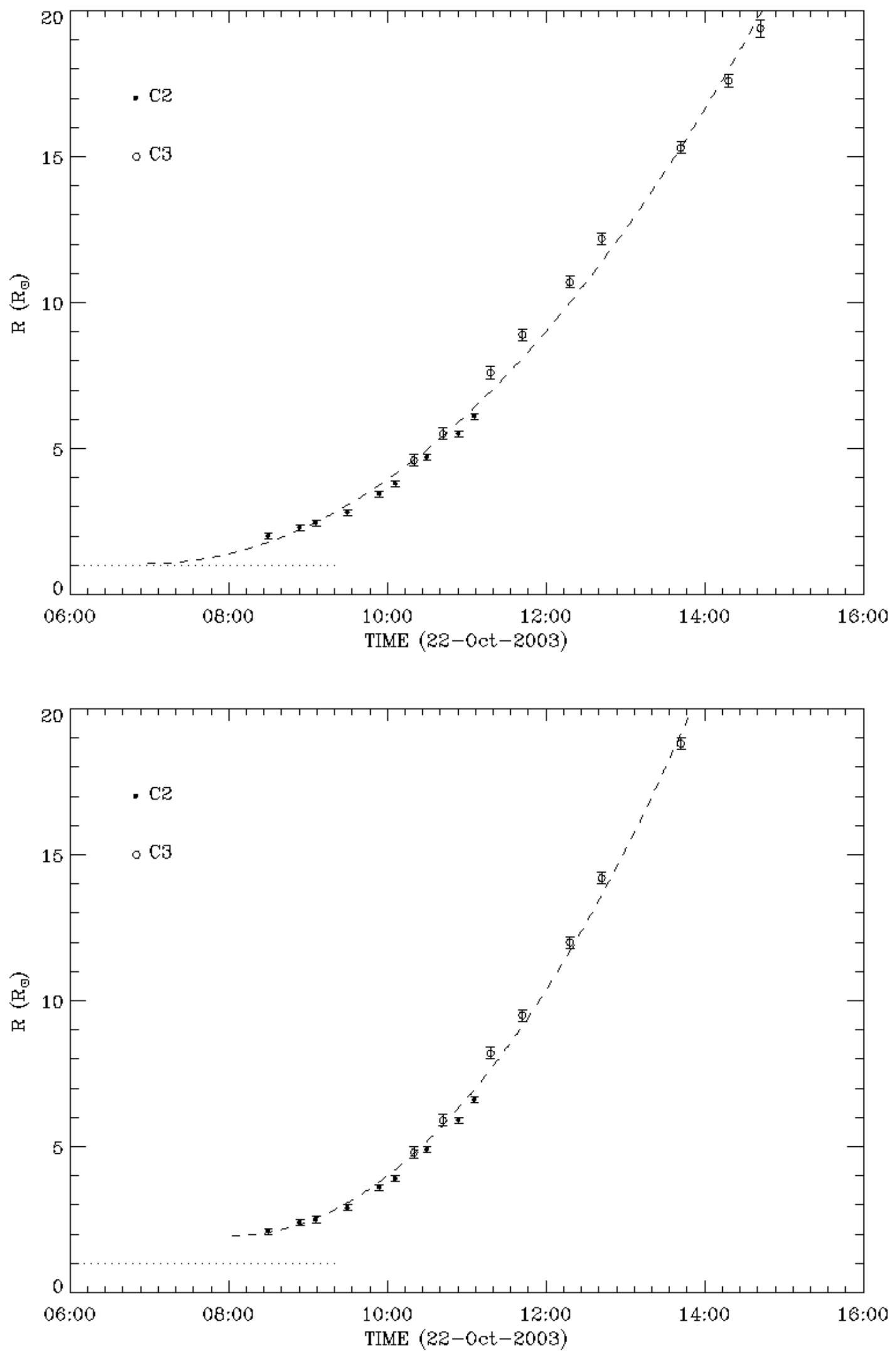

**Figure 8.** Height-time plots of the center $C$ of $\mathrm{CME}_{22}$ for the two solutions of propagation along the $OC$ direction, forward (upper panel) and backward (lower panel).

## 6. Mass of the CME of 22 October 2003

To determine the mass of $\mathrm{CME}_{22}$, we selected the C2 image obtained at 10:30 as the optimum case since the CME fully extends over the C2 field of view without truncation. The image was processed and photometrically calibrated as described in Section 2 except that the vignetting function (used as a radial filter) was not applied. Our method consisted in properly scaling the radiance of the corresponding synthetic images, that is determining the electron density

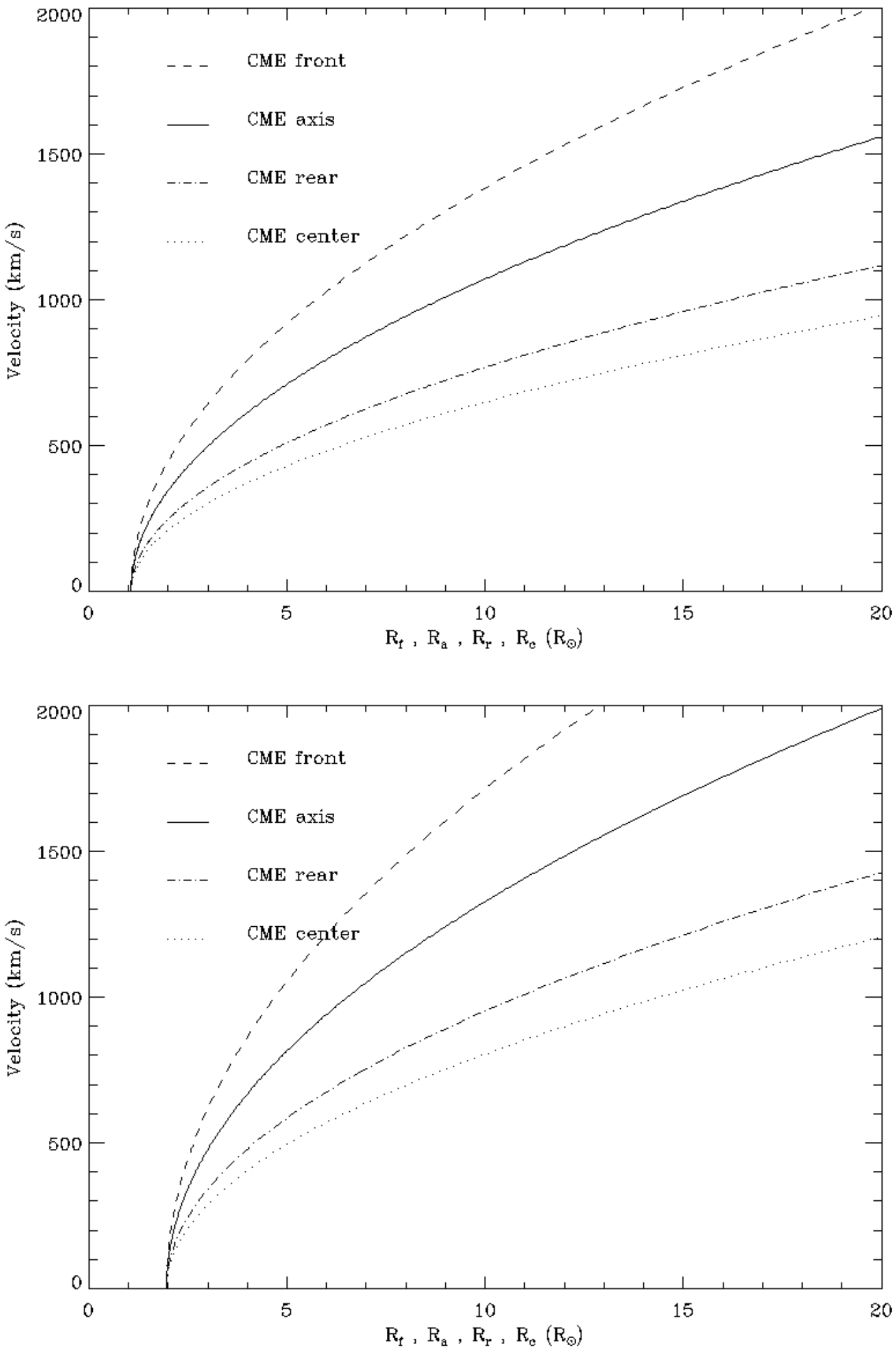

**Figure 9.** Velocity profiles of different parts of $CME_{22}$ for the two solutions of propagation, forward (upper panel) and backward (lower panel).

$N_{e,0}$ arbitrarily set in Section 3 and then, calculating the mass of the simulated CME. A binary mask was constructed and applied to both observed and synthetic images to precisely delimit the extent of $CME_{22}$ and block the circular occulted zone as well as remnants of the streamers present in the observed images (Figure 10). The ratio of the integrated radiances in the unmasked regions then defined the sought for scaling factor and in turn, the absolute value of $N_{e,0}$. Our method has the advantage of by-passing the complex internal structure of $CME_{22}$

**Table 2.** Parameters for the two solutions of propagation of the CME of 22 October 2003.

| Solution<br>Unit | Lat<br>° | Long<br>° | Ac<br>m s$^{-2}$ | Ar<br>m s$^{-2}$ | Aa<br>m s$^{-2}$ | Af<br>m s$^{-2}$ |
|---|---|---|---|---|---|---|
| Forward | $4 \pm 1$ | $5 \pm 1.5$ | 27 | 36 | 49 | 62 |
| Backward | $4 \pm 1$ | $175 \pm 1.5$ | 31 | 46 | 60 | 76 |

Lat and Long: latitude and longitude of CME source.
Ac: Acceleration of CME center.
Ar: Acceleration of CME rear.
Aa: Acceleration of CME axis.
Af: Acceleration of CME front.

**Table 3.** Characterization of the kinematics of the CME of 22 October 2003 from different catalogs.

| Catalog<br>Unit | ID | $V_1$<br>km s$^{-1}$ | $V_2$<br>km s$^{-1}$ | Accl<br>m s$^{-2}$ |
|---|---|---|---|---|
| ARTEMIS | CR2008-144 | 910 | - | - |
| CACTUS | 0028 | 578 | - | - |
| CDAW | 08:30:32 | 719 | 1080 | 37 |
| SEEDS | 222 | 285 | - | 10 |
| SEEDS | 223 | 477 | - | 77 |

ID: CME identification in the catalogs.
$V_1$: Linear speed.
$V_2$: 2nd order speed at $20\,R_\odot$.
Accl: Acceleration.

whose masses are ascribed to the shells of the APC model. The composition of $CME_{22}$ plasma was assumed to be fully ionized hydrogen plasma with 10% helium (Vourlidas2002) leading to a mass of $1.7\,10^{16}$ g.

Only two catalogs list CME masses and for $CME_{22}$, ARTEMIS gives $1.0\,10^{16}$ g and CDAW $1.1\,10^{16}$ g mentioning a poor mass estimate. The agreement is quite remarkable as the two methods are totally different: ARTEMIS uses synoptic maps whereas CDAW relies on the images themselves. These "routine" estimates are however not corrected for geometric effects. Our present larger value may then be explained by two combined effects: i) the de-projection effect that leads to an increase of the mass as illustrated for instance by Table 1 of Floyd and Lamy (2019) for 13 CMEs and ii) the full account of the CME volume offered by the APC model.

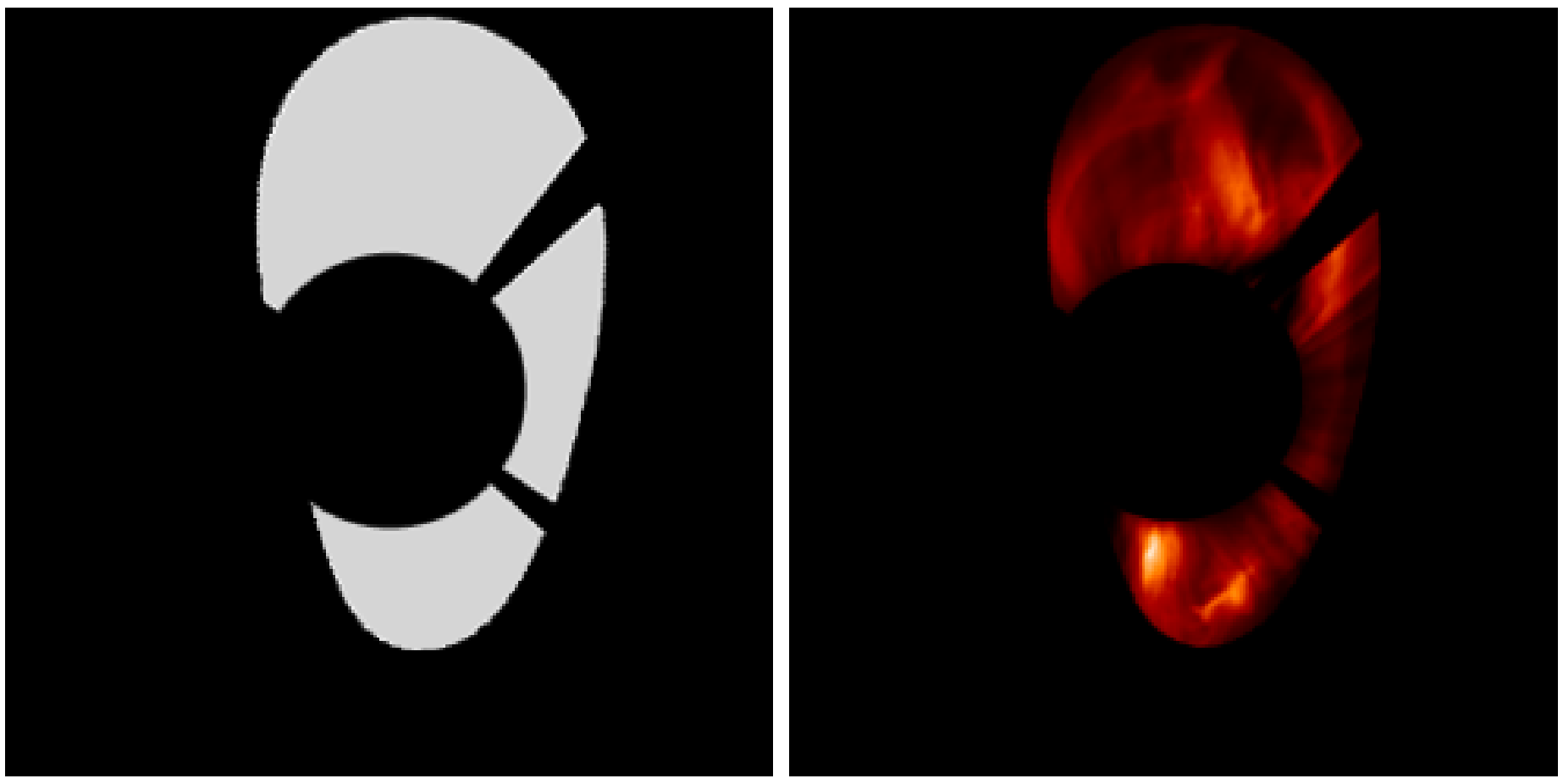

**Figure 10.** Image of the binary mask (left panel) and its application to the C2 image at 10:30 (right panel) For illustrative purpose, we display a filtered image of $\mathrm{CME}_{22}$, not the image used for the mass determination.

## 7. Source of the CME of 22 October 2003

Violent solar eruption occurred from 18 October to 8 November 2003 during which 80 CMEs were detected, most of them being fast, wide, and hence very energetics (Gopalswamy et al., 2005). The activity culminated in late October and early November 2003 with the most energetic events of Solar Cycle 23 and subsequently became known as the "Halloween event". Gopalswamy et al. (2005) suggested active region NOAA 10484 as the source of $\mathrm{CME}_{22}$ with however a question mark in their Table 1. AR 10484 was indeed very active on 22 October 2003 around the estimated launch time of 7:15 of $\mathrm{CME}_{22}$ as illustrated by the EUV observations performed by the Extreme-Ultraviolet Imaging Telescope (EIT: Delaboudinière et al. (1995) in the 195Å bandpass (Fe XII)). We document this activity by an animated sequence (supplementary material) constructed from images obtained from 06:12 to 08:24 after subtracting the first image, by a difference image and by a contemporary map of the solar disk from the Heliophysics Feature Catalogue[5] both displayed in Figure 11. The EIT observations reveal recurrent jets appearing on short time scales. Coronal dimming appeared around AR 10484, first northward starting at 6:24 and then southward starting at 08:36, and lasted more than 3 hr. Post-eruptive arcades started at 06:35 and evolved slowly until 12:00. Whereas this activity is likely at the origin of the CME (ejecta) which appeared some 20 minutes before $\mathrm{CME}_{22}$ propagating eastward as mentioned in Section 2, its connection to $\mathrm{CME}_{22}$ is not straightforward.

[5] https://voparis-helio.obspm.fr/hfc-gui/index.php

AR 10484 was at the east limb on 18 October and at a longitude of -20° on 22 October at 7:00 (measured westward from the central meridian) whereas the source of $\mathrm{CME}_{22}$ was at longitude +5° (our favored forward solution). The implied deflection of 25° is rather large and would have further taken place over a very short path followed by radial propagation as observed on the C2 images. Both evidences concur in seriously questioning the connection of $\mathrm{CME}_{22}$ to AR 10484 and led us to re-consider the backward solution. Figure 12 displays two EIT images obtained half a Carrington rotation before and after 22 October, that is on 8 October and 5 November, thus revealing the opposite solar hemisphere. They do not reveal any noticeable persistent activity that would have led to a CME. Both the difficulty with the backward propagation of the CME pointed out in Section 5 and the above fact led us to conclude that the backward solution is indeed unrealistic.

Finally, we examined two synoptic maps of the coronal radiance[6] at 3.5 $\mathrm{R}_\odot$ which extend over CR 2008 and CR 2009 (Figure 13). The CME activity prominently driven by the three active regions ARs 10484, 10486, and 10488 dominated in the eastern sector from 18 to 28 October, then faded and switched to the western sector thereafter. $\mathrm{CME}_{22}$ clearly stands out of this picture and appears totally uncorrelated with the flurry of CMEs pervading the eastern sector. $\mathrm{CME}_{22}$ therefore appears as an interloper in the set of 18 CMEs which have been unquestionably connected to AR 10484 by Gopalswamy et al. (2005) and its source remains a puzzle, especially since it cannot be considered as a stealth CME known to be often slow and diffuse.

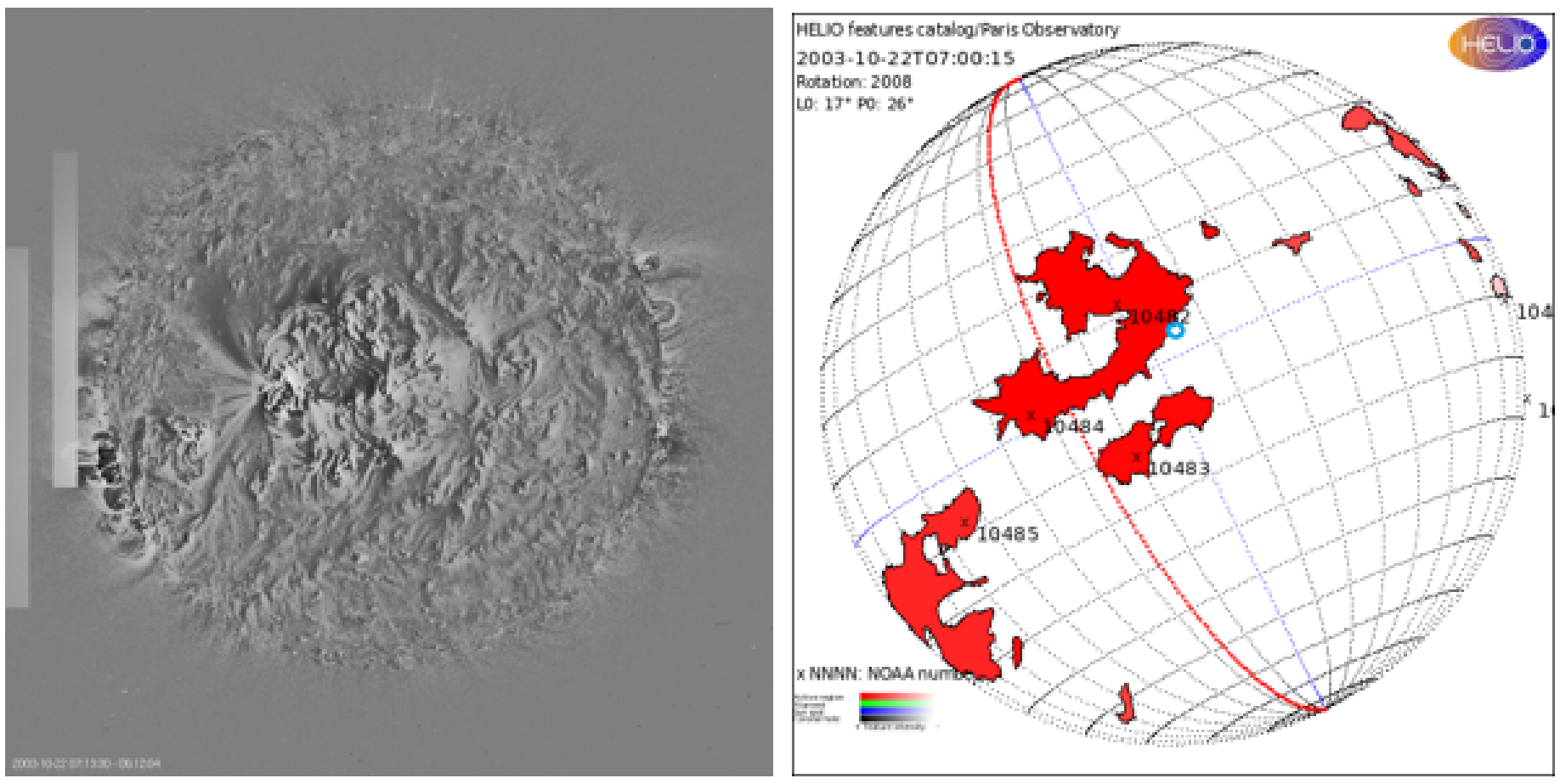

.

**Figure 11.** Left panel: Difference of two EIT image taken on 22 October 2003 at 07:13:30 and at 06:12:04 in the 195Å bandpass showing the coronal dimming associated with eruptive activity around around AR 10484. Right panel: Contemporary map of the solar disk from the Heliophysics Feature Catalogue on 22 October 2003 at 07:00:15 showing the extent of the active regions labeled by their NOAA numbers and indicated by crosses. The blue circle indicates our determination of the location of the source of $\mathrm{CME}_{22}$

[6]http://idoc-lasco-c2-archive.ias.u-psud.fr

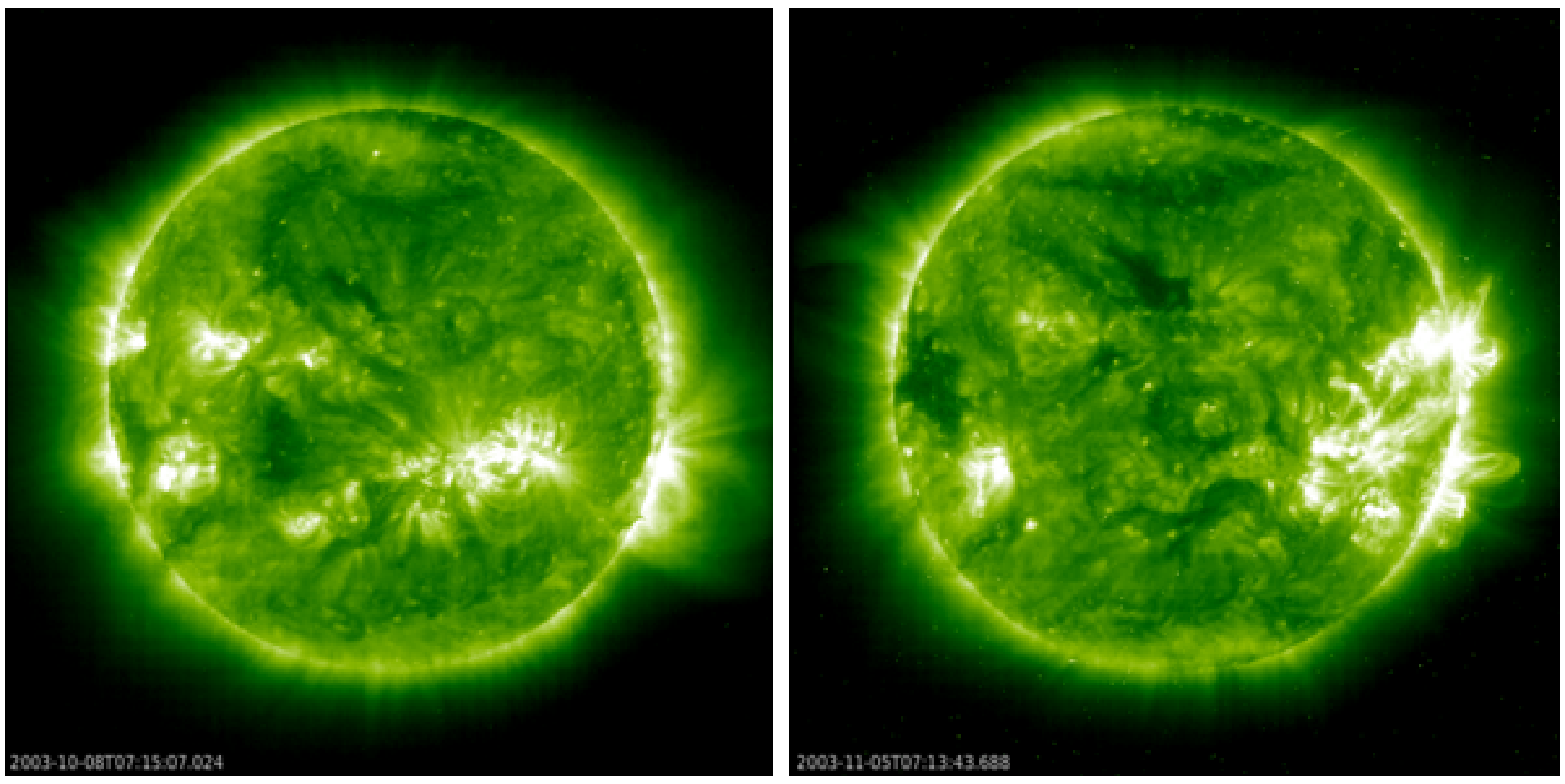

**Figure 12.** EIT images taken on 8 October (left panel) and on 5 November 2003 in the 195Å bandpass.

## 8. Discussion

To our knowledge, this is the first time that the shape and the propagation of a CME can be reconstructed based on observations from a single vantage point and furthermore, that an asymmetric configuration is required. The shape of $\mathrm{CME}_{22}$ is best represented by an asymmetric plasma cloud, but is probably an asymmetric magnetic flux rope whose legs remain unseen due to a quasi-halo geometry.

Self-similar expansion of CMEs in the corona is very common as established in early works, for instance Low (1984). More recently and based on a data set of 475 CMEs selected for their clear flux rope signatures, Balmaceda et al. (2020) found that 65% of them exhibit a self-similar evolution at $10\,\mathrm{R}_\odot$, reaching 70% at $15\,\mathrm{R}_\odot$. $\mathrm{CME}_{22}$ is therefore well in line with this widespread behavior. In contrast, radial propagation is much less common and deflections due to the influence of the background coronal magnetic and flow pattern were noted in early works, for instance MacQueen, Hundhausen, and Conover (1986) and confirmed by many subsequent articles, for instance Cremades and Bothmer (2005). Likewise and for similar reasons, rotations of the magnetic axis of the flux rope – and even more complex distortions – are often observed (e.g. Thernisien, Howard, and Vourlidas (2006); Krall et al. (2006)). However, in their investigation of 54 magnetic cloud (MC) and ejecta (EJ) CMEs, Kim, Park, and Moon (2013) found that the majority of the MC-associated CMEs are ejected along the radial direction (whereas many of the EJ-associated CMEs are not) with a mean speed of 946 km $\mathrm{s}^{-1}$, both results highly consistent with our findings. It is likely that fast, energetic CMEs are less prone to suffer deviations and distortions than slow CMEs. In addition, $\mathrm{CME}_{22}$ probably benefited from a rather quiet coronal environment as indicated by the synoptic maps shown in Figure 13, thus favoring its radial propagation and its fixed orientation.

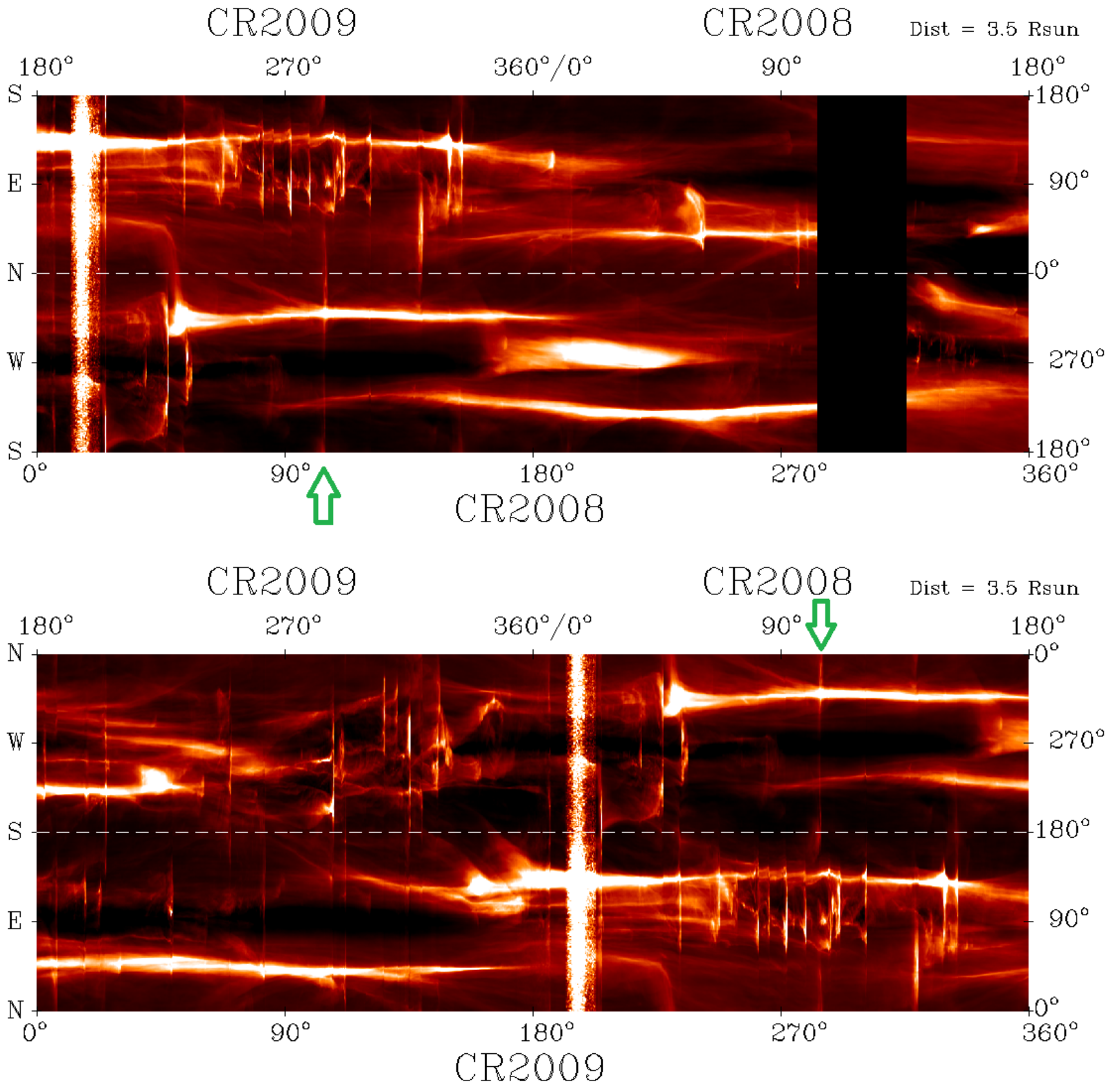

**Figure 13.** Two synoptic maps of the corona at 3.5 $R_\odot$ displaying simultaneously the east and west hemispheres illustrate the CME activity during CR 2008 and CR 2009. Time runs from right to left and Carrington longitude increases from left to right with two different origins according to the hemisphere. The two maps differ in their centering, either north or south. This is intended to offer unsplit views of the north and south hemispheres. All CMEs appear as vertical streaks and $CME_{22}$ is indicated by the two green arrows. Its large north lobe is best seen on the upper map whereas its smaller south lobe is best seen on the lower map. The wide, bright, fuzzy band results from images overwhelmed by impacts of energetic particles on the C2 CCD detector, a "snow storm" caused by the impressive halo CME of 28 October 2003.

The front of $CME_{22}$, that is its fastest part, reached a speed of ≈ 1000 km $s^{-1}$ at the edge of the C2 field of view and of ≈ 2000 km $s^{-1}$ at 20 $R_\odot$. This ranks $CME_{22}$ in the class of the rather rare high speed CMEs. Indeed, according to the cumulative distribution of apparent speeds over Solar Cycles 23 and 24, only ≈5 % of CMEs have speeds exceeding 1000 km $s^{-1}$in C2 field of view as shown by Figure 35 of Lamy et al. (2019). $CME_{22}$ maintained a constant acceleration throughout this region with values ranging from 27 m $s^{-2}$ for its center to 62 m $s^{-2}$ for its front. Although there are pronounced differences in the distributions of the apparent accelerations reported by different catalogs as shown by Figures 37 and 39 of Lamy et al. (2019), the above values appear rather extreme. Whereas it is generally admitted that slow CMEs tend to accelerate and to the contrary, fast CMEs tend to decelerate while crossing the C2 and C3 fields of view (Webb and Howard, 2012), this was not the case of $CME_{22}$. Its mass of $1.7\,10^{16}$ g here again places $CME_{22}$ among the largest values of the distribution functions as given in Figure 12 of Lamy et al. (2019). Combined with a speed of the front of 1000 km $s^{-1}$ at the edge of the C2 field of view this yields a kinetic energy of $8.4\,10^{31}$ erg placing $CME_{22}$ among the most energetic CMEs

It remains challenging to identify the source of $CME_{22}$. On the one hand, a putative connection with AR 10484 raises the difficulty of a strong deflection over a very short path before its radial propagation. Furthermore, it would be difficult to understand why 18 CMEs clearly associated with this active region, particularly that one which appeared some 20 minutes before $CME_{22}$, were seen in the eastern equatorial sector whereas $CME_{22}$ appeared totally outside this region. On the other hand, the backward extrapolation of its trajectory to the solar surface does not point to any active region.

## 9. Conclusion

In this article, we have presented an investigation of a CME tracked throughout the LASCO-C2 and C3 fields of view on 23 October 2003 during an episode of extreme solar activity. We highlight below our most significant results.

- Thanks to favorable geometric conditions, the shape and the propagation of the CME could be reconstructed by forward modeling based on observations from a single vantage point, a unique achievement.
- Its shape was best represented by an asymmetric plasma cloud, but was probably an asymmetric flux rope whose legs were concealed by the coronagraph occultation and by the body of the CME itself.
- Its propagation was characterized by radial propagation and self-similar expansion with no detectable rotation of its axis.
- Its kinematics stands out by extreme speeds ($\approx 1000$ km $s^{-1}$ at the edge of the C2 field of view and $\approx 2000$ km $s^{-1}$ at $20\,R_{\odot}$) and constant accelerations (up to 62 m $s^{-2}$ ) of its different parts. Combining with its large mass of $1.7\,10^{16}$ g propels the CME to the class of the most energetic events.
- Its source remains unclear and could not be unambiguously connected to AR 10484. It rather appears as an interloper in the set of 18 CMEs which have been unquestionably connected to this active region.

**Acknowledgments** The LASCO-C2 project at the Laboratoire Atmosphères, Milieux et Observations Spatiales is funded by the Centre National d'Etudes Spatiales (CNES). LASCO was built by a consortium of the Naval Research Laboratory, USA, the Laboratoire d'Astrophysique de Marseille (formerly Laboratoire d'Astronomie Spatiale), France, the Max-Planck-Institut für Sonnensystemforschung (formerly Max Planck Institute für Aeronomie), Germany, and the School of Physics and Astronomy, University of Birmingham, UK. SOHO is a project of international cooperation between ESA and NASA.

**Disclosure of Potential Conflicts of Interest** The authors declare that they have no conflicts of interest.